\documentclass[conference]{IEEEtran}
\IEEEoverridecommandlockouts
\usepackage{cite}
\usepackage{amsmath,amssymb,amsfonts}
\usepackage{algorithmic}
\usepackage{amsmath}
\usepackage{graphicx}
\usepackage{booktabs}
\usepackage{textcomp}
\usepackage{float}
\usepackage{xcolor}
\def\BibTeX{{\rm B\kern-.05em{\sc i\kern-.025em b}\kern-.08em
    T\kern-.1667em\lower.7ex\hbox{E}\kern-.125emX}}
    
\begin{document}

\title{A Comprehensive Survey on Feature Extraction Techniques Using I/Q Imbalance in RFFI}

\author{
\IEEEauthorblockN{Muhammad Aqib Khan}
\IEEEauthorblockA{
School of Computation, Information and Technology\\
\textit{Technical University of Munich}, Munich, Germany\\
aqib.khan@tum.de}
\and
\IEEEauthorblockN{Muhammad Usman Siddiqui}
\IEEEauthorblockA{School of Science and Electrical Engineering\\
\textit{Habib University}, Karachi, Pakistan\\
ms05996@alumni.habib.edu.pk}
}

\maketitle

\begin{abstract}
The proliferation of Internet of Things (IoT) devices has intensified the need for secure authentication. Although traditional encryption-based solutions can be robust, they often impose high computational and energy overhead on resource-limited IoT nodes. As an alternative, radio frequency fingerprint identification (RFFI) exploits hardware-induced imperfections—such as Inphase/Quadrature (I/Q) imbalance—in Radio Frequency (RF) front-end components as unique identifiers that are inherently difficult to clone or spoof. Despite recent advances, significant challenges remain in standardizing feature extraction methods, maintaining high accuracy across diverse environments, and efficiently handling large-scale IoT deployments. This paper addresses these gaps by offering a comprehensive review of feature extraction techniques that harness I/Q imbalance for RFFI. We also discuss other hardware-based RF fingerprinting sources, including power amplifier nonlinearity and oscillator imperfections, and we survey modern machine learning (ML) and deep learning (DL) approaches that enhance device identification performance.
\end{abstract}

\begin{IEEEkeywords}
I/Q Imbalance, Feature Extraction, Fingerprinting, RFFI
\end{IEEEkeywords}

\section{Introduction}
With the explosive growth of IoT devices, the need for secure and efficient ways to verify device authenticity has become paramount. While traditional cryptographic methods (e.g., symmetric key encryption) protect data confidentiality by converting plaintext into ciphertext, they do not inherently validate the legitimacy of the transmitting device. To address the challenge of distinguishing genuine devices from imposter or rogue transmitters, researchers have explored radio frequency fingerprint identification (RFFI). By leveraging hardware-induced imperfections in RF front-end components, RFFI provides unique, device-specific signatures that are extremely difficult to clone \cite{b1}. As such, it offers a complementary layer of security focused on ensuring only recognized devices can participate in network communication. 

This approach, known as radio frequency fingerprint identification (RFFI), uses device-specific characteristics introduced during manufacturing. Early RFFI implementations relied primarily on manually engineered features, which required extensive domain expertise yet yielded limited accuracy and robustness. With the advent of deep learning (DL), these limitations have been mitigated through the automatic extraction of highly discriminative features directly from raw I/Q data. In particular, convolutional neural networks (CNNs) and other DL architectures have shown promise in isolating device-specific impairments such as I/Q imbalance \cite{b1, b2}.

I/Q imbalance is especially significant in direct-conversion receivers, where the in-phase (I) and quadrature (Q) components should be orthogonal and of equal amplitude. In practice, imperfections such as Direct Current (DC) bias, amplitude mismatch, and phase offset cause notable deviations from these ideal conditions \cite{b2}. Substantial research has therefore been devoted to compensating for I/Q imbalance, with proposed techniques spanning both time-domain and frequency-domain approaches \cite{b3, b4}. Beyond compensation, emerging studies highlight that I/Q imbalance can serve as a unique signature for device identification, especially when combined with advanced ML or signal processing methods. Indeed, recent investigations into unsupervised contrastive learning and federated learning have showcased the potential of I/Q imbalance for improving RFFI systems in scenarios with limited data \cite{b1, b3}.

Despite these promising findings, challenges remain in achieving standardized feature extraction, ensuring high accuracy in dynamic or hostile environments, and efficiently scaling to massive IoT networks. This paper provides a comprehensive review of I/Q imbalance-based RFFI feature extraction methods. We additionally discuss other notable RF hardware impairments—such as power amplifier nonlinearity and oscillator imperfections—and examine ML/DL-driven solutions that bolster device identification.  

The rest of this paper is organized as follows: Section~II introduces the fundamentals of I/Q imbalance, including its origins and consequences for RF systems. Section~III delves into existing feature extraction techniques based on I/Q imbalance, highlighting strengths, weaknesses, and relevant case studies. Finally, Section~IV concludes the paper and discusses prospective research directions for RFFI.  

\section{Understanding I/Q Imbalance}
I/Q imbalance arises from hardware imperfections in mixers, ADC/DAC converters, or filters, causing mismatches in amplitude and phase between the in-phase (I) and quadrature (Q) components of a signal. Ideally, these components are of equal amplitude and $90^\circ$ out of phase, but deviations distort the baseband signal and constellation diagram \cite{b2}.

The received baseband signal with I/Q imbalance can be modeled as:
\begin{equation}
r(t) = \alpha x(t) + \beta x^*(t),
\label{eq:I/Q_imbalance_model}
\end{equation}
where \(x(t)\) is the ideal signal, \(x^*(t)\) its complex conjugate, and \(\alpha, \beta\) are coefficients capturing imbalance:
\begin{equation}
\alpha = \cos \theta + j \varepsilon \sin \theta, \quad 
\beta = \varepsilon \cos \theta + j \sin \theta,
\end{equation}
with \(\varepsilon\) and \(\theta\) denoting gain and phase mismatches, respectively \cite{b15}. These parameters shift the constellation points from their ideal positions.

\begin{figure}[ht]
    \centering
    \includegraphics[width=0.48\textwidth]{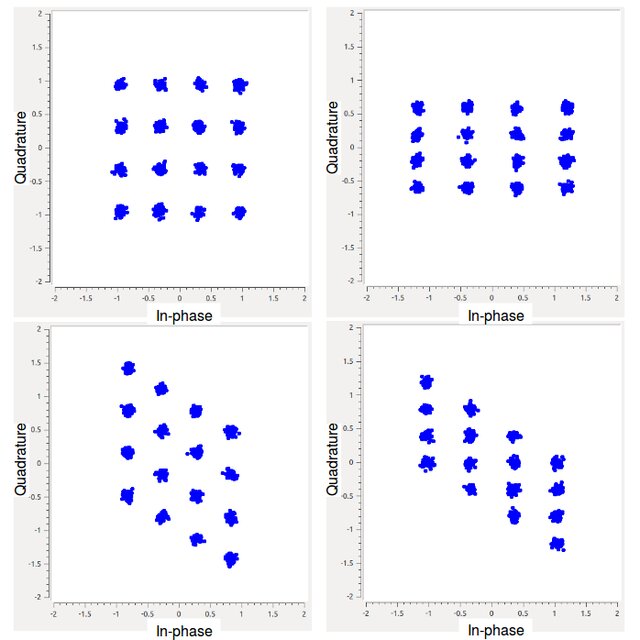}
    \caption{Examples of I/Q imbalance effects on a 16QAM constellation (SNR = 20\,dB). Top Left: no mismatch; Top Right: phase mismatch = $30^\circ$; Bottom Left: gain mismatch = 0.9; Bottom Right: both mismatches. Adapted from \cite{Wong2019}.}
    \label{fig:I/Q_imbalance_constellation}
\end{figure}

Figure~\ref{fig:I/Q_imbalance_constellation} demonstrates how I/Q imbalance affects constellations:
\begin{itemize}
    \item \textbf{No Imbalance}: Symmetrical 16-QAM (Quadrature amplitude modulation) points.
    \item \textbf{Phase Imbalance}: Rotational skew in phase.
    \item \textbf{Gain Imbalance}: Elliptical distortion.
    \item \textbf{Both}: Combined severe distortion.
\end{itemize}

\begin{figure}[ht]
    \centering
    \includegraphics[width=0.48\textwidth]{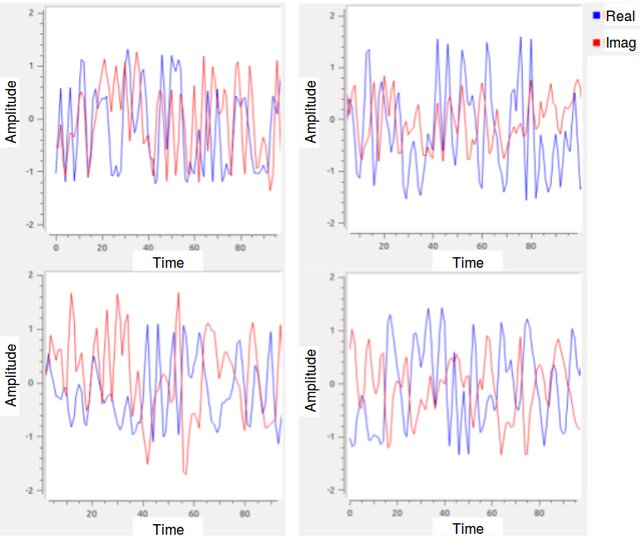}
    \caption{Time-domain effects of I/Q imbalance on a 16QAM signal (SNR = 20\,dB). Adapted from \cite{Wong2019}.}
    \label{fig:I/Q_imbalance_time_domain}
\end{figure}

In the time domain (Fig.~\ref{fig:I/Q_imbalance_time_domain}), I/Q imbalance disrupts the ideal $90^\circ$ phase shift between I and Q signals, visibly distorting waveforms. While degrading metrics like error vector magnitude (EVM) and demodulation accuracy, these distortions create unique signatures. Leveraging this property, I/Q imbalance aids Radio Frequency Fingerprint Identification (RFFI), enabling secure device authentication in IoT applications.

\section{Feature Extraction/Estimation Techniques Using I/Q Imbalance}

\subsection{Adaptive Filter-Based Feature Extraction}
Wang et al.\ propose a method for estimating I/Q imbalance that leverages adaptive filtering and is particularly targeted at LTE-RACH (Random Access Channel) signals~\cite{b16}. Their framework begins by modeling the received baseband signal (including I/Q imbalance) as:
\begin{equation}
\begin{split}
y(t) & = 2\bigl[y_I(t) + j\,y_Q(t)\bigr] \\
& = \bigl[x_I(t) + j(1 + \varepsilon)e^{j\phi}x_Q(t)\bigr] \otimes h(t) + \omega(t)
\end{split}
\label{eq:imbalance_model}
\end{equation}
where $\varepsilon$ denotes gain imbalance, $\phi$ is the phase imbalance, and $h(t)$ is the channel response.

Time synchronization and frequency offset compensation are performed first, followed by channel estimation using a Least Mean Square (LMS)-based adaptive filter:
\begin{equation}
h(n+1) = h(n) + \mu\,e^{*}(n)\,x(n),
\label{eq:lms_filter}
\end{equation}
where the I/Q imbalance parameter $\mu$ is calculated via conjugate correlation:
\begin{equation}
\mu = \frac{1 + (1 + \varepsilon)e^{j\phi}}{2}.
\label{eq:mu_extraction}
\end{equation}
Achieving 96.01\% as top accuracy when tested on LTE mobile devices and Universal Software Radio Peripheral (USRP) platforms, the method strongly depends on precise synchronization and offset compensation, making it well-suited to controlled scenarios.

\begin{figure}[h]
\centering
\includegraphics[width=0.45\textwidth]{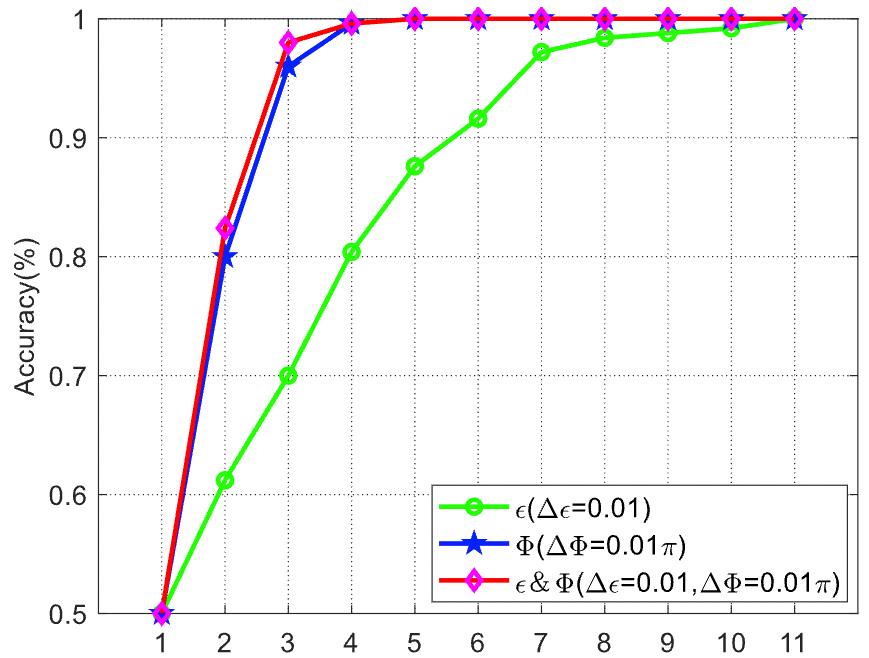}
\caption{Performance illustration of the Adaptive Filter Technique~\cite{b16}}
\label{fig:adaptive_filter}
\end{figure}

\begin{table*}[h]
\centering
\footnotesize
\caption{Comparison of Techniques}
\label{tab:comparison}
\begin{tabular}{|p{1.7cm}|p{1.7cm}|p{2.4cm}|p{1.3cm}|p{1.3cm}|p{1.5cm}|p{1.5cm}|p{2.0cm}|}
\toprule
\textbf{Technique} 
  & \textbf{Performance}
  & \textbf{Testing Criteria}
  & \textbf{Channel Dep.}
  & \textbf{Feature Type}
  & \textbf{Comp. Power}
  & \textbf{Mem. Reqs.}
  & \textbf{Resources} \\
\midrule

\textbf{Adaptive Filter} 
  & 96.01\% (LTE vs.\ USRPs) 
  & Tested on LTE-based systems using Zadoff–Chu (ZC) sequences, validated under varying channel conditions 
  & Yes 
  & Amplitude/ Phase Distortion 
  & Moderate 
  & Moderate 
  & Precise synchronization (ZC-based) \\
  
\textbf{Channel-Correlation} 
  & Over 90\% 
  & Cross-scenario validation using DMRS and CP signals; shows robustness to environmental variations 
  & No 
  & Higher-order Statistics 
  & High 
  & High 
  & Minimal hardware constraints (relies on DMRS) \\

\textbf{Signal Space Representation} 
  & Over 90\% (SNR = 15dB) 
  & Simulated using 5 analog transmitters with I/Q imbalance; tested on 400 signals (train/test split: 50\%) 
  & Yes 
  & Amplitude/ Phase Distortion 
  & Moderate 
  & Low 
  & No demodulation required; applicable to analog and digital modulation \\
  
\bottomrule
\end{tabular}
\end{table*}
\subsection{Channel-Correlation Based Feature Extraction}
\begin{figure}[h]
\centering
\includegraphics[width=0.45\textwidth]{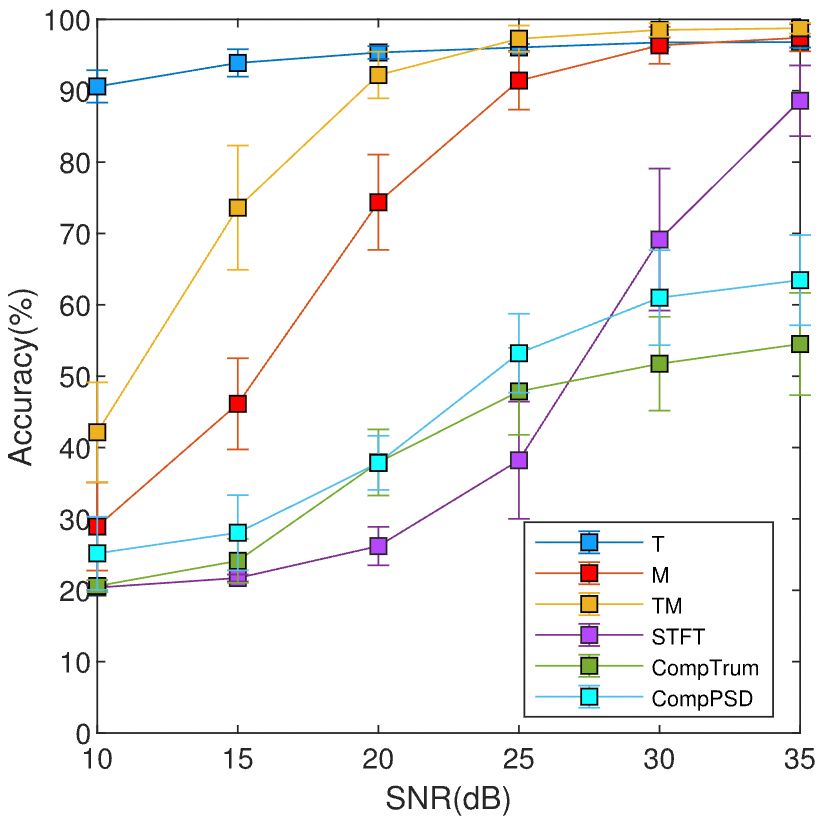}
\caption{Performance illustration of the Channel-correlation Technique~\cite{b17}}
\label{fig:channel_correlation}
\end{figure}

Peng et al.\ present a technique that mitigates channel effects by exploiting the strong correlation among adjacent subcarriers in cellular systems~\cite{b17}. The approach combines demodulation reference signal (DMRS) analysis with cyclic prefix (CP) examination. First, DMRS signals are used to extract steady-state features:
\begin{equation}
\mathrm{RFFl\_rs1}(\lambda) = \frac{\sum |Y_{\mathrm{rs1}}[k]|}{\sum |Y_{\mathrm{rs1}}[k]|},
\label{eq:dmrs}
\end{equation}
while transient-on features are gathered from CP analysis:
\begin{equation}
\mathrm{RFF1}(\lambda) = \frac{ \bigl|\sum \hat{y}_1[n]\,\hat{y}_1^{*}[n + N]\bigr|}{\lambda}.
\label{eq:cp_analysis}
\end{equation}
By leveraging both time-domain (transient) and frequency-domain (steady-state) information, as well as channel-robust differential features, the method maintains a high (92.95\%) accuracy in dynamic environments with significant variations. 
as shown in figure \ref{fig:channel_correlation}.

\subsection{Signal Space Representation-Based Feature Extraction}
Zhuo et al. presents a feature extraction approach leveraging the signal space representation of I/Q imbalance which begins by modeling the I/Q modulator with gain and phase imbalance parameters. The in-phase (I) and quadrature (Q) components are described by $x_I(t)$ and $x_Q(t)$, respectively, where $x_I(t)$ represents the baseband signal and $x_Q(t)$ may be its Hilbert transform for analog signals or baseband signal for digital modulation  \cite{b15}. The received signal is polluted by white Gaussian noise and can be mathematically expressed as:

\begin{equation}
    r(t) = x(t) + x^*(t) + v(t) \label{eq:r_t}
\end{equation}
where $x(t)$ is the transmitted signal, $x^*(t)$ is its complex conjugate, and $v(t)$ is the additive noise. By combining the received signal with its conjugate, the signal space representation is obtained:
\begin{equation}
    r(t)^H = AX + V
    \label{eq:hermitian}
\end{equation}
where A is the matrix that captures the I/Q imbalance parameters, X is the transmitted signal, and V is the noise. The autocorrelation matrix of the signal is expressed as:

\begin{equation}
    R_Y = A R_X A^H + R_V
    \label{eq:resulting_equation}
\end{equation}

where $R_X$ represents the autocorrelation of the transmitted signal and $R_V$ represents the noise power. Based on this representation, the signal-to-noise ratio (SNR) is estimated using eigenvalue decomposition of the autocorrelation matrix:
\begin{equation}
    \text{SNR} = \frac{\sigma_s^2}{\sigma_v^2} 
    \label{eq:snr}
\end{equation}

where $\sigma_s^2$ and $\sigma_v^2$ are the power of the signal and noise, respectively. The extracted fingerprint features are constructed as:

\begin{equation}
    \text{Feature} = \begin{bmatrix}
    \text{Re}(R_Y) \\
    \text{Im}(R_Y)
    \end{bmatrix}
    \label{eq:feature_matrix}
\end{equation}

which encode the I/Q imbalance distortions and serve as unique identifiers for specific emitters
The proposed methodology was evaluated using simulation experiments involving five analog transmitters, each with distinct I/Q imbalance parameters. Each transmitter generated 400 signals, with 50\% used for training and the remaining 50\% for testing. The method was compared against two existing feature extraction techniques: bispectrum-based and Hilbert-Huang transform-based methods. The experimental results demonstrated that the proposed method outperforms these techniques, particularly in terms of classification accuracy, as shown in Figure 5 of the referenced study. At an SNR of 15 dB, the features extracted using the proposed approach exhibited superior clustering capabilities, facilitating accurate differentiation of transmitters. The method’s performance is robust, achieving higher recognition rates with fewer sampled points compared to competing methods, which require extensive sampling for accurate bispectrum or time-frequency energy distributions.

\begin{figure}[h]
    \centering
    \includegraphics[width=0.95\linewidth]{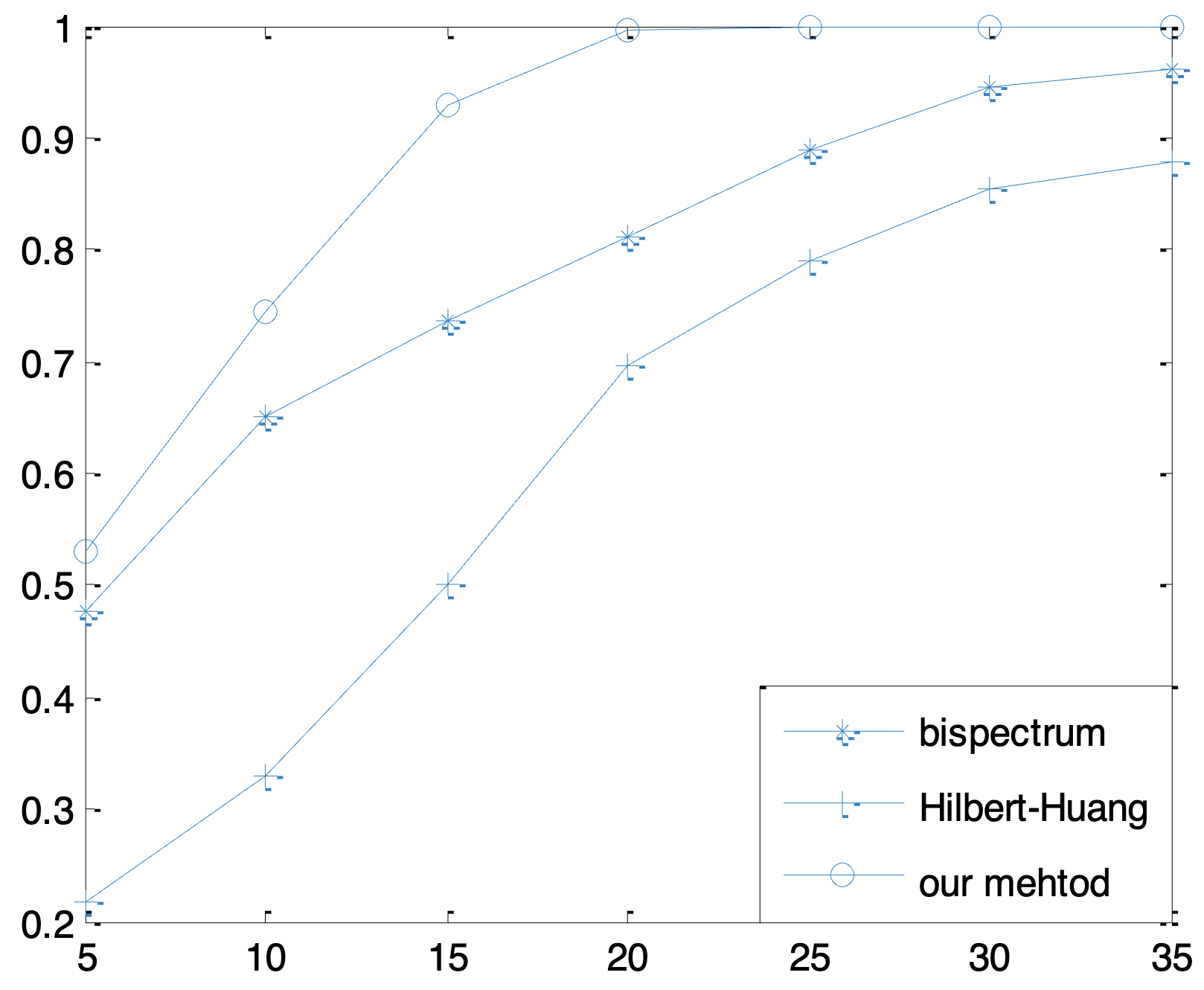}
    \caption{Performance illustration of the Wavelet Denoising Technique\cite{b15}}
    \label{fig:signal}
\end{figure}

Table \ref{tab:comparison} contrasts these aforementioned techniques in terms of performance, testing scenarios, channel dependency, and resource usage. Each method presents distinct advantages, such as high accuracy under controlled conditions or robustness across diverse environments, and the choice of approach should reflect application requirements (e.g., computational budget, synchronization constraints, or channel variability).

The Signal Space Representation technique offers a robust approach for both analog and digital modulation schemes without requiring demodulation. It is highly effective in identifying emitters under moderate SNR conditions (e.g., 15 dB) and demonstrates low memory requirements, making it suitable for resource-constrained systems. Adaptive filters are highly accurate for synchronized signals but rely heavily on hardware-dependent Carrier Frequency Offset (CFO) compensation. Channel correlation techniques are robust across dynamic and static environments, offering a hybrid approach that balances transient and steady-state features.

Researchers must select the most suitable method based on specific application requirements. For resource-constrained systems or scenarios involving both analog and digital signals, the Signal Space Representation method is an optimal choice due to its simplicity, noise robustness, and low memory overhead. Adaptive filters are ideal for controlled laboratory settings with well-synchronized signals, where high accuracy is paramount despite hardware dependencies. For applications involving dynamic and multipath-rich environments, the channel-correlation technique remains the most suitable, as it effectively handles transient and steady-state features, albeit with higher computational and memory demands.

\section{Other Sources for RFFI—Hardware Imperfections}
Beyond I/Q imbalance, several other RF hardware imperfections can serve as valuable sources for device fingerprinting \cite{b8}. These hardware-level variations arise from manufacturing processes or component aging and can be harnessed as reliable identifiers.

\subsection{Power Amplifier Characteristics} 
Power amplifiers exhibit unique nonlinearities, which yield discriminative features such as amplitude compression, phase distortion, and memory effects \cite{b10}. Machine and deep learning models can capture these PA-specific traits, improving classification performance under practical, real-world conditions.

\subsection{Oscillator-Based Features}
All oscillators introduce variability in both frequency and timing domains, including:
\begin{itemize}
   \item Carrier frequency offset deviations
   \item Phase noise characteristics
   \item Clock skew and sample timing jitter
\end{itemize}
These oscillator imperfections can significantly enhance device discrimination, particularly for devices with otherwise similar RF profiles \cite{b9, b12}.

\subsection{Front-end Component Features}
Other analog front-end components, such as filters or matching circuits, contribute further to the overall RF signature:
\begin{itemize}
   \item Filter response shifts due to component tolerances
   \item Amplifier bias point differences
   \item Impedance mismatches
\end{itemize}
Modern ML approaches are well-suited to uncovering these subtle, multifaceted variations \cite{b13, b14}.

\subsection{Composite Hardware Signatures}
Recent research demonstrates that aggregating multiple hardware-level features—from I/Q imbalance, PA characteristics, and oscillator deviations—can notably increase identification accuracy \cite{b11}. Deep learning architectures that automatically learn and fuse these signatures show particular promise for robust device classification across fluctuating environmental or channel conditions.
\section{Conclusion}
This paper has surveyed the primary feature extraction techniques that exploit I/Q imbalance for radio frequency fingerprint identification (RFFI), as well as additional hardware-based RF fingerprinting sources. Although compensation methods exist to mitigate I/Q imbalance in communications, leveraging residual imperfections for device identification continues to demonstrate increasing viability, particularly in IoT environments where energy constraints can limit calibration efforts.

Adaptive filter-based methods, channel-correlation approaches, and signal space representation techniques each present unique advantages and trade-offs in terms of performance, channel dependency, and computational demands. The signal space representation approach, in particular, offers a novel framework for feature extraction that is applicable to both analog and digital modulation schemes, eliminates the need for demodulation, and performs well under moderate SNR conditions with low memory requirements. Furthermore, power amplifier nonlinearity, oscillator imperfections, and other front-end variances offer complementary or alternative sources of fingerprinting data. Modern ML and DL techniques can efficiently extract and combine these hardware-specific features, creating robust, high-accuracy identification systems suitable for large-scale IoT deployments.

In the future, establishing standardized feature extraction frameworks and evaluating performance in more diverse and dynamic environments will be crucial for widespread adoption. Additionally, scalable ML/DL solutions that can adapt to varying device populations and environmental conditions remain an active area of research. The integration of techniques such as signal space representation with advanced ML/DL approaches holds significant potential for improving the accuracy and robustness of RFFI systems in complex real-world scenarios.`

\bibliographystyle{IEEEtran}

\end{document}